\documentclass[fleqn,10pt]{wlscirep}
\title{Centrifugal photovoltaic and photogalvanic effects driven by structured light}
\usepackage{balance}
\usepackage{graphicx}
\usepackage{epsfig}
\usepackage{color}
\author[1] {J. W\"atzel}
\author[1] {J. Berakdar}

\affil[1]{Institut f\"{u}r Physik, Martin-Luther-Universit\"{a}t Halle-Wittenberg, 06099 Halle, Germany}

\begin{abstract}
Much efforts  are devoted to  material structuring  in a quest to enhance the photovoltaic effect.
We show that structuring light in a way it transfers orbital angular momentum to  semiconductor-based
 rings results in a steady charge accumulation  at the outer  boundaries that be utilized for the generation
 of an open circuit voltage  or a
 photogalvanic (bulk photovoltaic) type current.
 This effect which stems both from  structuring  light and matter (confinement potentials), can be magnified even at fixed moderate intensities, by increasing the orbital angular momentum of light which strengthens the effective centrifugal potential that repels  the charge outwards.
Based on  a full numerical time propagation of the carriers wave functions in the presence of light pulses we demonstrate how the   charge buildup  leads to a useable voltage or directed  photocurrent whose amplitudes and directions are controllable by the light pulse parameters.
\end{abstract}

\begin{document}
\flushbottom
\maketitle
\thispagestyle{empty}
\noindent
\section*{Introduction}
The feasibility of light carrying orbital angular momentum (OAM) \cite{Allen1992,ref3,ref4,ref5,ref6,ref7,Allen} opened  the way
for exciting  new applications  ranging from electronics  and life sciences to  quantum information, astronomy, or optical telecommunications\cite{Terriza2007,mair2001entanglement,barreiro2008beating,boyd2011quantum,padgett2011tweezers,furhapter2005spiral,key-16,
torres2011twisted,andrews2011structured,foo2005optical,he1995optical,wang2008creation,hell2007far}. For instance,  OAM beams allow to trap, rotate and manipulate microscopic objects \cite{Allen2,TL_Barreiro,FrieseNat1998}, atoms, molecules \cite{TL_Romero,TL_Awfi,TL_Araoka} as well as Bose-Einstein condensates\cite{Helmerson_Torres2011twisted}. An OAM beam may also drive electric current loops in quantum rings with an associated local, light-controlled magnetic field pulses\cite{TL_Quinteiro2009,quinteiro2011orbital}.
The phase front associated with   OAM beam forms a helical shape. Thus in cylindrical coordinates  with $z$ direction being along the light propagation,
the field spatial distribution contains
a term $\exp(i\ell_{\rm OAM}\varphi)$.  Here $\varphi$ is the azimuthal angle and $\ell_{\rm OAM}$ is the topological charge of the optical vortex.
  Allan \emph{et al.}\cite{Allen1992} showed that helical beams  (realized for example as Laguerre-Gaussian (LG) modes) carry OAM with respect to $z$   direction, the amount of which is  $\ell_{\rm OAM}\hbar$ per photon.\\
 Numerous techniques are available for generating OAM beams: They can be created from usual light sources\cite{Allen1992,Heckenberg1992,Kennedy2002}, other approaches include computer-generated holograms screened on a spatial light modulator (SLM)\cite{TL_Alicia,TL_Curtis,TL_Ostrovsky}, astigmatic mode converters\cite{ref3}, spiral phase plates \cite{Beijersbergen1994}, and conversion of spin angular momentum to OAM in inhomogeneous anisotropic plates\cite{marrucci2006optical}. These methods have different strengths and limitations. In general beam generation in connection with SLM has a low efficiency and the overall beam quality is restricted by the pixel size of the nematic liquid crystal cells. The other techniques are static and therefore cannot be  controlled dynamically. A newer approach for generating and manipulating OAM beams is realized with a ring resonator based geometry\cite{schulz2013integrated}. Optical vortices with radii independent on the topological charge can be generated based on the width-pulse approximation
of Bessel functions\cite{ostrovsky2013generation,garcia2014simple}. \\
A key element of OAM light when interacting with matter is the time change of the carriers' OAM. This  implies a  torque  exerted on the charge carriers\cite{FrieseNat1998,ONeil2002,Simpson97,Gahagan96,babiker1994light,andrews2012angular} rendering so qualitatively new ways to steer the orbital motion by light. For instance as demonstrated in Ref.\,\cite{TL_JW}, an electronic wave packet in a semiconductor stripe irradiated with an OAM light spot acquires a transverse drift whose direction and amplitude is governed by the parameters of the OAM beam.\\
Here we explore a further effect of an OAM beam focused on a micro sized  GaAs-AlGaAs-based quantum ring\cite{Ring_Levy,Ring_Mailly} causing intra conduction band transitions which is shown to result in a centrifugal drift of the carriers and thus to a time-sustainable charge imbalance between the inner and out ring boundaries.  This charge separation (mimics an intraband photovoltaic effect) is exploitable for the generation of an open circuit voltage which can be tuned in magnitude by increasing the light topological charge at a fixed frequency and intensity, i.e. without additional heating.\\
In a next step we demonstrate a structured light-induced  photogalvanic-type (or bulk photovoltaic-type) mechanism. Such
 an effect occurs conventionally for non-structured light  in media with noncentrosymmetric crystal structure \cite{galva1,galva2}. Typical examples
  are  doped lithium niobate or bismuth ferrite. We suggest and show by full-fledge numerical simulations
   that the proposed optical vortex-induced photogalvanic effect is  systematically  controllable by changing the properties of light and by appropriate nano structuring of the system. A possible realization are quantum rings \cite{Ring_Levy,Ring_Mailly} with spiral phase plates (SPP)\cite{Beijersbergen1994} deposited atop. A Gaussian light
   ray traversing SPP  acquires OAM that is subsequently delivered to the ring carriers. Combined with the quantum confinement effect, the rings act as a light-driven charge wheel enhancing the current in an attached wire. The potential of the current predictions are endorsed by recent experiments \cite{kikawa} on n-doped bulk GaAs irradiated with OAM pulses.
    It was shown experimentally that the sample attains indeed  an orbital angular momentum. The theory proposals presented here point in addition to the potential of nanostructuring the sample and exploiting the quantum confinement effects in addition to the structure of the light wave fronts.

\section*{Results}
%\subsection{Interaction of nano structures with structured light}
%
\begin{figure}[!t]
\centering
    \includegraphics[width=.85\textwidth]{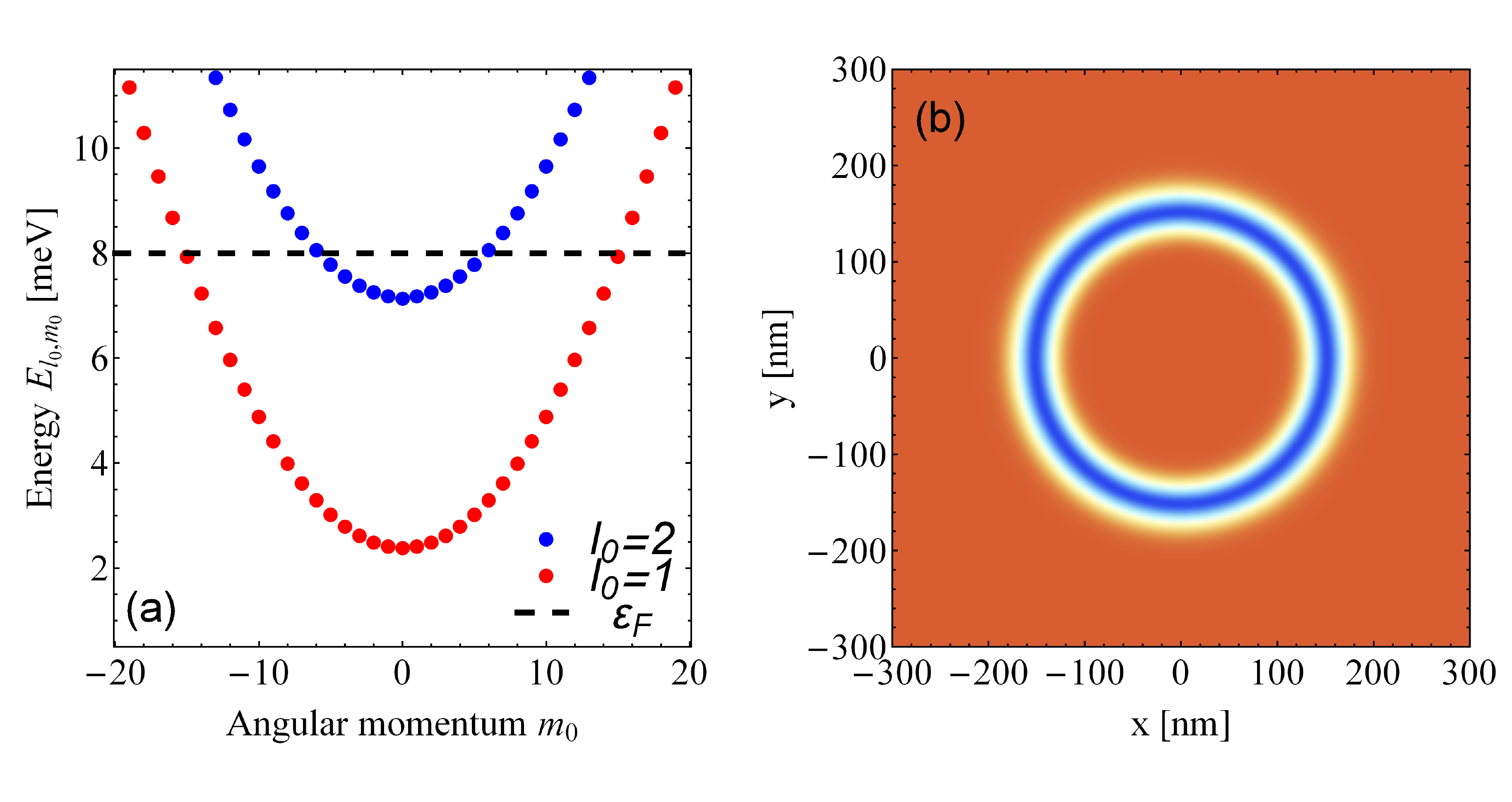}
\caption{(a) Energy levels $E_{l_0,m_0}$ of the ring structure. The Fermi energy $\varepsilon_F$ is marked by the dashed horizontal line. (b) Initial LDOS of the considered 2D-system.}
\label{fig:init}
\end{figure}
We employ   experimentally feasible OAM laser beams impinging vertically onto  a ballistic GaAs-AlGaAs-based nano-size ring such as those reported in Refs.\,\cite{Ring_Levy,Ring_Mailly}. We investigate the  intraband quantum dynamics of the conduction band and calculate the transient and steady-state  time evolution of the charge. To utilize the aforementioned centrifugal photovoltaic effect for direct current
generation, we study in a further setup the OAM-laser driven charge current in wires attached to the ring. The thickness of the ring (and later of the wires) is small such that no dynamic occurs along the $z$-direction due  to
quantum size effects, and thus one may safely restrict the considerations to  the $xy$-plane in which the ring structure is embedded.
 We assume a uniform effective mass $m^*=0.067m_e$ and an average ring radius  $r_0=150$\,nm with a width $\Delta r=50$\,nm and a Fermi energy $\varepsilon_F=8$\,meV.  With these predetermined values the electronic structure of the electrons in the
 conduction band of the considered rings \cite{Ring_Levy,Ring_Mailly} is well captured. We note that the frequency and the intensity of the
 light are chosen such that only intraband dynamics in the conduction band is triggered.
  In fig.1 %~\ref{figf:init}
  the numerically calculated stationary, unperturbed  subbands relevant for our study, and the local density of states (LDOS) are shown.
An initial energy level $E_{l_0,m_0}$ is classified according to the quantum numbers $l_0$ and $m_0$,
where $l_0=1, 2, 3, ...$ characterize the radial motion in the ring.  The angular motion is quantified by the angular momentum $m_0$.
As expected, the initial states are degenerated with respect to the clock-wise and anti-clock-wise angular motion,   i.e. $E_{l_0,m_0}=E_{l_0,-m_0}$
and hence the system is current-less. Furthermore, the radial density distribution is angularly homogeneous and is radially symmetric  with respect to  $r_0$, as also demonstrated below, meaning that there is no voltage drop between the inner and outer ring boundaries.
In what follows we will be interested in non-invasive excitations near the Fermi energy in which case the
independent effective single particle picture is still viable\cite{WCTanModel1996,Imry,Ring2,Ring3,Alex1}.\\
Applying  a weak  monochromatic laser pulse  carrying orbital angular momentum we trigger the time propagation of the single-particle wave functions $\Psi_{l_0,m_0}(x,y,t)$ that evolves from the stationary state labeled with the quantum numbers $l_0$ and $m_0$ at the time $t=0$.
Technically we obtain $\Psi_{l_0,m_0}(x,y,t)$  by solving fully numerically for the time-dependent  Schr\"{o}dinger equation in the presence of the confining potential  and the spatially inhomogeneous laser vector potential $\vec{A}(x,y,t)$ with frequency $\omega$ and amplitude $A_0$.
 The OAM beam is propagating along the $z$-direction and is focused  vertically on the ring.
The light has a right circular polarization, i.e.  the polarization vector  is $\vec{e}$ is $\vec{e}_-=\sqrt{1/2}(\vec{e}_x-i\vec{e}_y)$.
Below we choose  $\ell_{\rm OAM}=-10$ and
an amplitude corresponding to a peak intensity of $I_{\rm TL}=10^6~$W/m$^2$. The photon energy is $\hbar\omega=5$~meV ($\lambda=247\,\mu$m) and
the pulse duration is characterized by two optical cycles, i.e. $\tau=1.65$\,ps. The beam waist is chosen in a way that the radial intensity profile is not larger than $200$~nm, i.e. $w_0=55$~nm.  As discussed below, a possible experimental realization is to  deposit  on the ring of interest an appropriate spiral phase plate and irradiating the whole structure by a focused Gaussian beam. The transmitted light is converted into OAM light resulting in the electromotive effects presented below. Aside from this idea, it is worthwhile
 to mention  recent achievements in the development of  metamaterial-based lenses allowing for strong  focusing\cite{zhang2008superlenses,zhao2012hyperlens}. The key point of such a lens arrangement is that the spatial profile of the electric field in the focused light spot may be modified but the corresponding topological charge is conserved.\\
As discussed in\cite{FrieseNat1998,ONeil2002,Simpson97,Gahagan96,babiker1994light,andrews2012angular}
 OAM  beam transfers  its OAM  when interacting with a dielectric particle. This change in OAM causes a  torque. The total torque, within the paraxial approximation, can be given by the photon flux multiplied by the total angular momentum of the beam.
 In our case of LG modes the  amount of transferrable angular momentum is given by $(\ell_{\rm OAM}+\sigma_z)\hbar$, where $\sigma_z$ is the helicity of the circularly polarized light (in our case $\sigma_z=-1$). This same amount also applies to the vortex-induced break in the clockwise- anticlockwise symmetry, meaning that for large $\left|\ell_{\rm OAM}\right|$ large charge currents are achievable. We note that the
 "torque" associated with this OAM change of the carriers has its origin not only in the vector potential of the OAM-LG beam but
 equally important in  the confinement potential that hinders the charge density to escape.
\begin{figure}[!t]
\centering
\includegraphics[width=.85\textwidth]{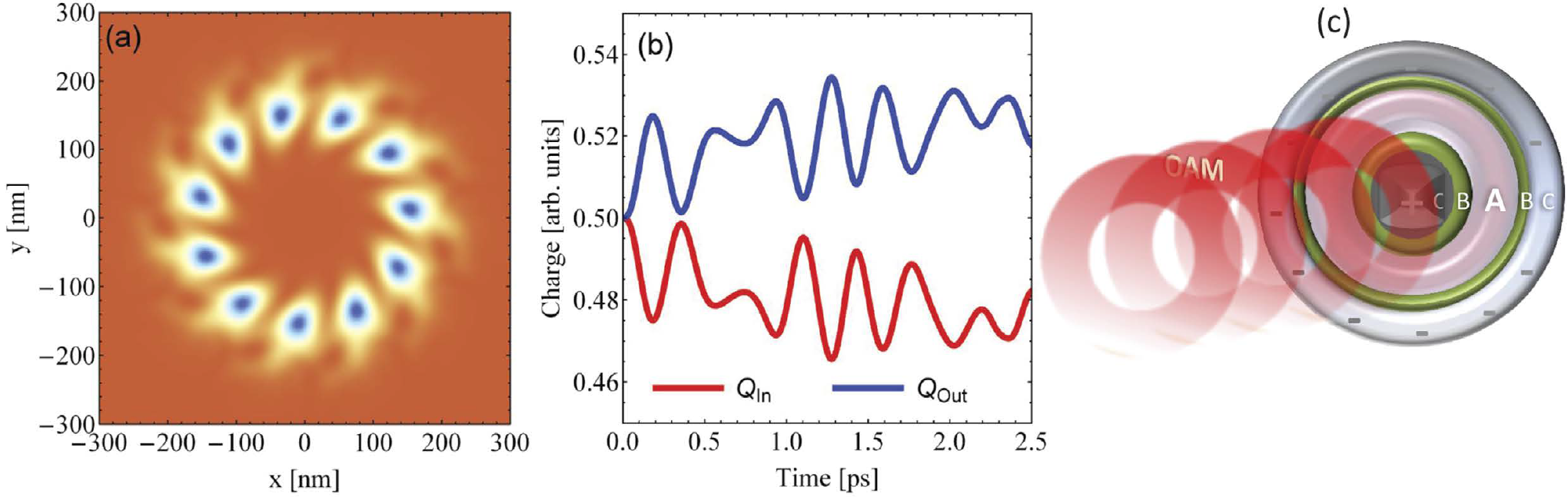}
%\caption{
%\label{fig:Dens}
%\end{figure}
%%
%%
%\begin{figure}
%\centering
%\includegraphics[width=12cm]{fig2c.eps}[!t]
\caption{
(a) LDOS of the system after a propagation time of $t=2$~ps. (b) Inner and outer density $Q^{\rm In/Out}$ in dependence of the propagation time. (c)
a possible scheme for an OAM-driven open circuit voltage generation. OAM the ring (A) is separated by two thin tunneling barriers (B) from two electrodes (C).}
\label{fig:voltage}
\end{figure}
A demonstration is depicted in fig.2(a) % \ref{fig:Dens}
where  the ring local density of states (LDOS) is shown at a time $t=2$~ps, which means after the laser pulse. We clearly  notice that the initial left-right radial symmetry with respect to $r_0$ as well as the clockwise and anticlockwise angular symmetries are
broken hinting so on the appearance of a radial  charge accumulation at the outer ring boundaries (due to the enhancement in the effective centrifugal potential) and the emergence of a charge current loop. We notice 11 nodal angular structures which are  explainable by selection rules (i.e., conservation of angular momentum) and considering that we are exciting initially completely symmetric (degenerate) states with a circular polarized OAM beam with $\ell_{\rm OAM}=-10$, i.e. the total amount of angular momentum transferred to the ring structure is $-11\hbar$. The direction of the observed whirl is invertible by changing the sign of the topological charge (not shown for brevity). A scheme to collect this charge imbalance as a vortex-driven open circuite voltage is illustrated in fig.2(c). How the driven charge may tunnel the boundaries of the rings is shown below.\\

\section*{Discussion}

Fig.2(a) % \ref{fig:Dens}
evidences a charge density drift of the initial equilibrium state to outer radii over the course of the application time
of the OAM beam. This is due to the enhanced  radially repulsive centrifugal force upon an effective increase in the angular momentum by
$\ell_{\rm OAM}$ which means that this photovoltaic effect can be enlarged by tuning $\ell_{\rm OAM}$ (as long as the centrifugal potential
does not overcome the confinement leading so to electron emission).
To quantify this observation we calculate the charge density in the inner and outer area of the ring structure corresponding to an initial state with the quantum numbers $l_0$ and $m_0$ as
%\begin{equation}
$Q_{l_0,m_0}^{\rm In}(t)=\int_0^{2\pi}{\rm d}\varphi\int_0^{r_0}{\rm d}r\,r |\Psi_{l_0,m_0}\left(r,\varphi\right)|^2$ %\end{equation}
 for the inner area,  and
%\begin{equation}
$Q_{l_0,m_0}^{\rm Out}(t)=\int_0^{2\pi}{\rm d}\varphi\int_{r_0}^\infty{\rm d}r\,r |\Psi_{l_0,m_0}\left(r,\varphi\right)|^2$
%\end{equation}
for the outer ring area.
For the whole conduction subbands these  quantities are found as
\begin{equation}
Q^{\rm In/Out}(t)=\sum_{l_0,m_0}f(l_0,m_0,t)Q_{l_0,m_0}^{\rm In/Out}(t).
\label{Eq:Qinout}
\end{equation}
In Eq.~(\ref{Eq:Qinout}) $f(l_0,m_0,t)$ stands for the non-equilibrium distribution function. The relaxation processes (electron-phonon scattering, simultaneous scattering by impurities and phonons or electron-electron scattering) are introduced phenomenologically by means of a single (averaged) quantity, the relaxation time $\tau_{\rm rel}$. The non-equilibrium distribution function $f(l_0,m_0,t)$ is evaluated within the relaxation time approximation by solving the Boltzmann equation\cite{ZimanBook} (we recall that we are considering  relatively weak, low-energy   excitations around
$\varepsilon_F$)
\begin{equation}
\frac{\partial f(l_0,m_0,t)}{\partial t}=-\frac{f(l_0,m_0,t)-f^0(l_0,m_0,\varepsilon_F)}{\tau_{\rm rel}}.
\label{Eq:NonEq}
\end{equation}
\begin{figure}[!t]
\centering
\includegraphics[width=.85\textwidth]{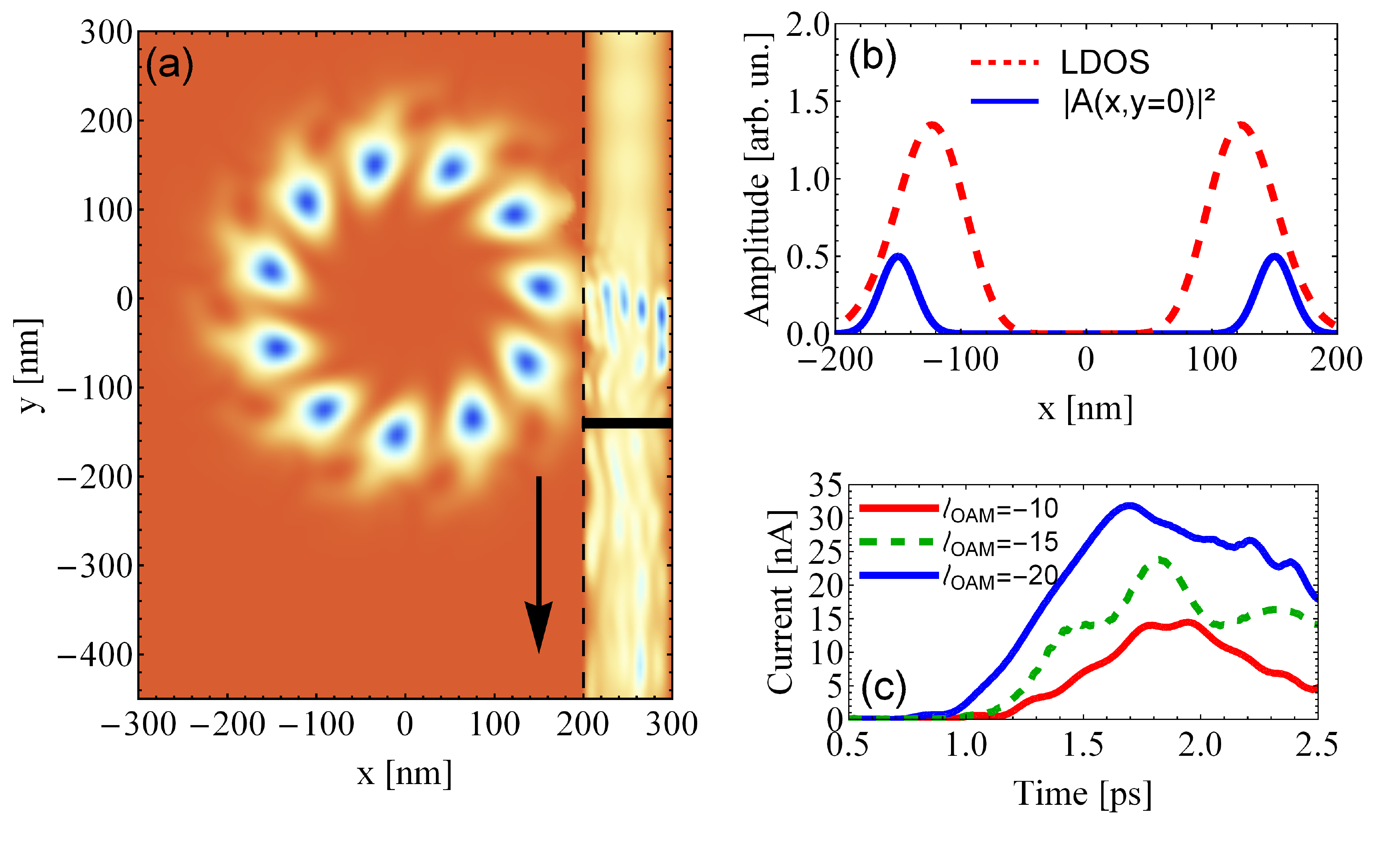}
\caption{(a) LDOS of the ring  attached to a conducting bar of the same material after a propagation time of $t=2.4$~ps. The detector is marked by the thick horizontal line, while the vertical dashed line shows the wire potential boundaries.
The arrow indicates  the direction of the flux charge density. (b)
the initial ring LDOS  with the OAM intensity profile (which does not touch the wire). (c) Time-dependent current through the detector at $y_d=-140$~nm for different topological charges $\ell_{\rm OAM}$. These currents are due to tunneling of the ring current density (tunneling of orbital moment and conversion into directed current flux density).}
\label{fig:Cur}
\end{figure}
The Fermi-Dirac distribution is
%\begin{equation}
$f^0(l_0,m_0,\varepsilon_F)=1/[ 1+ \exp\left( (E_{l_0,m_0}- \varepsilon_F)/k_BT \right) ] $
% \end{equation}
for a given temperature $T$ and Fermi energy $\varepsilon_F$ corresponding to the equilibrium. An averaged relaxation time of 25~ps is assumed\cite{Alex1} at a
constant Fermi energy.
 %is constant over the course of time by assuming that the number of particles is conserved.
 The evolution of the energy of the particle that develops from the initial stationary state $l_0,m_0$ can be obtained by calculating the time dependent matrix elements $E_{l_0,m_0}(t)=i\hbar\langle\Psi_{l_0,m_0}(t)|\frac{\partial}{\partial t}|\Psi_{l_0,m_0}(t)\rangle$ from which
 we infer  the levels that are involved in the process.
In fig.2(b) %~\ref{fig:Dens}(b)
the time dependence of $Q^{\rm In}$ and $Q^{\rm Out}$ are depicted.
At a time $t=0$ both quantities are equal which reflects the radial symmetry of the confinement potential around $r_0$.
Over the course of time the charge is redistributed (cf. fig.2(a)) in a way that the density is pressed to outer radii by the vortex beam.
At times well below the relaxation time, the evolution is unitary. The  frequencies of oscillations  exhibited in  fig.2(a) are
readily explained by the  frequencies of  the OAM-selection-rules-allowed transitions between levels near $\varepsilon_F$.  The generic long time behavior at finite temperatures might be to a certain extent be inferred from our previous study
 Ref.\,\cite{moskalenko2006revivals} on current relaxation in similar rings. There, it was shown that the current relaxation  (related
 to the population dynamics) is due to longitudinal acoustic phonons. The recent experiment \cite{kikawa} on OAM-excited transients in n-doped GaAs seems to indicate a long-lived component pointing to a possible OAM dependence of the electron-phonon coupling constant. 

Now we wish to extract the charge accumulation in the ring for an useable directed current.
To this end we wire the ring to a conductive straight channel at one side (cf. Fig.3).
This channel is not affected by the OAM beam, i.e. the light is
focused only on the ring and the carriers are allowed to tunnel to the wire. Such a potential landscape has already been realized
experimentally and can be modified at will by appropriate gating\cite{fuhrer2001energy,lorke2000spectroscopy}.
Theoretically, we need to change the potential landscape such that the carries inhibit the wire and the ring equally (we assume the ring and the wire are made of the same material at the same chemical potential).
\begin{figure}[!t]
\centering
\includegraphics[width=.85\textwidth]{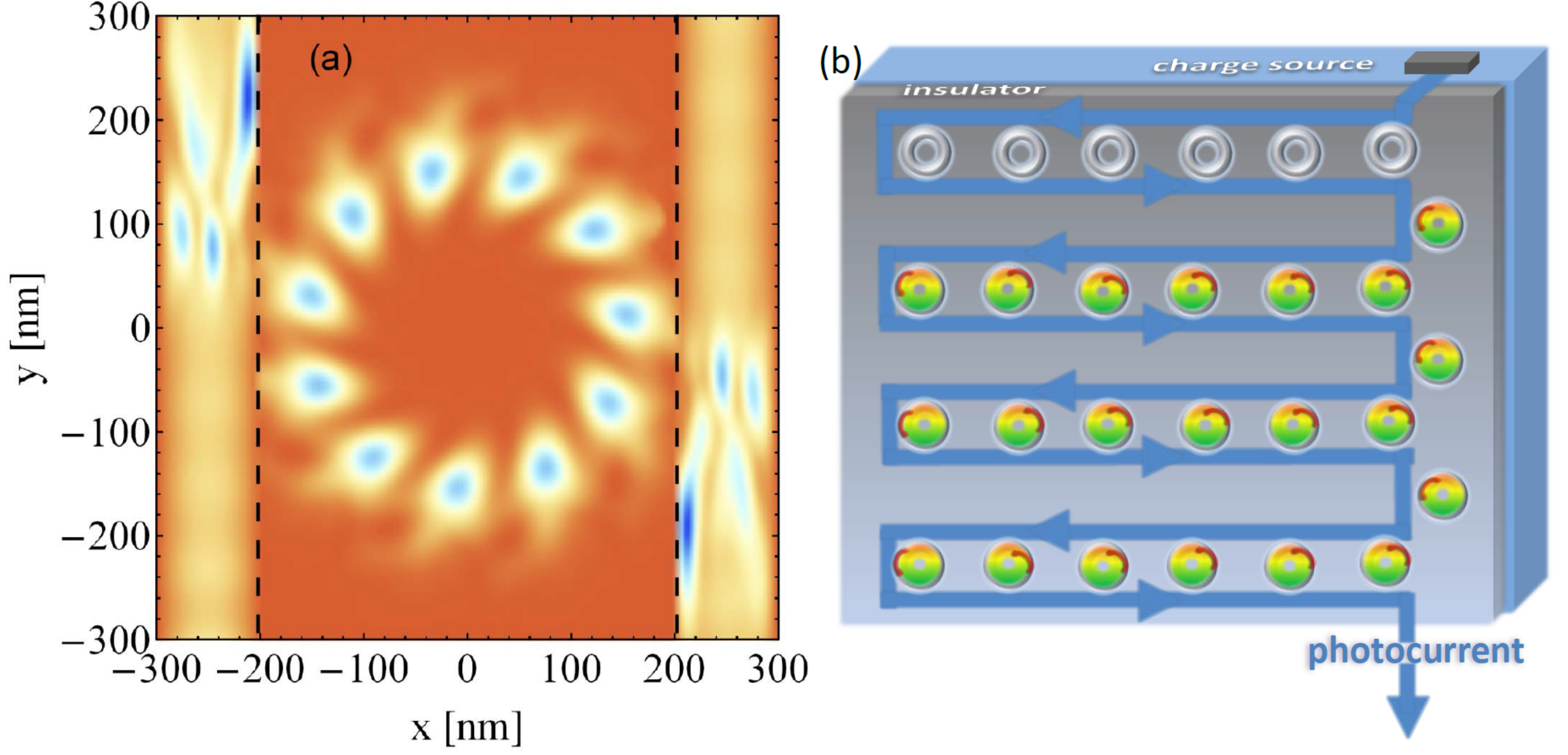}
\caption{
(a) LDOS of the ring connected to two wires while interacting with the OAM light with a topological charge $\ell_{\rm OAM}=-10$ at the time
$t=1.3$~ps. The pulse applied at $t=0$ has otherwise the same parameters as in fig.3. The currents in the wires are driven in opposite directions depending on the sign of the $\ell_{\rm OAM}$.
(b)
A schematic proposal for vortex-driven charge-wheels generating and controlling a photogalvanic-type current. On each mesoscopic ring (gray ring) a spiral phase plate is deposited that converts an incident non-structured light into OAM-carrying light wave (colored rings). As demonstrated explicitly in Fig.4a, this injects a directed current in coupled wire (blue line).}
\label{fig:2barb}
\end{figure}
Technically, the confining potential is such that the ring radius is $150$~nm, i.e. it is characterized for $x\leq200$\,nm by\cite{WCTanModel1996}. At 200\,nm a 100\,nm wide wire, i.e. $V(x>200\,{\rm nm},y)=0$, is attached (cf. fig.3). Taking into account a width $\Delta r=50$\,nm the effective barrier region between the quantum ring and the wire is around $25$\,nm wide. Since the modified potential $V(x,y)$ has no radial symmetry we characterize  the numerically calculated single-particle states by  the quantum numbers $n$ with the energy $E_n$. The calculation of these state reveals that the shape of the LDOS  in the ring region, i.e. $\sum_n f^0(n)|\Psi_n(x<200~{\rm nm},y,t=0)|^2$ is not dramatically different to the case without the conductive bar (in the equilibrium stateanyway). Fig.3 illustrates nicely the action of the OAM (acting solely on the ring): the ring charge density acquires a twist and a radial drift and tunnels to the wire crashing at the wire outer boundary (at $x=300$~nm). Due to the internal twist the reflected and the incoming waves in the wire form an interference pattern with an asymmetric density distribution with respect to the center of the ring. In fact this pattern is current carrying. The charge density flows mainly in negative $y$-direction after reaching the conductor, which reflects the sign  of the topological charge $\ell_{\rm OAM}$. We carefully checked the symmetry, i.e. that a change of the polarization direction and the sign of the topological charge leads to a flow of density in positive $y$-direction.
Quantitatively, the time-dependent current associated with a single particle state with the quantum number $n$ we obtain by calculating the probability current density  in $y$-direction as
\begin{equation}
j^y_{n}(x,y,t)=-\frac{1}{m*}{\rm Re}\left\{ \Psi^*_{n}(x,y,t)\left[i\hbar\partial_y + eA_y(x,y,t)\right]\Psi_{n}(x,y,t)\right\}.
\end{equation}
We position a detector in the conducting bar at $y_d=-140$~nm and calculate the time-dependent current through this detector with the help of
\begin{equation}
I_{n}(t)=\int_{x_1}^{x_2}{\rm d}x\,j^y_{n}(x,y_d,t).
\end{equation}
The bounds of the integration are the borders of the conductor at $x_1=200$~nm and $x_2=300$~nm. The detector is marked by the black horizontal line in fig.3(a). The total current of the system calculated as the weighted sum over all contributions of the partial currents $I_{n}(t)$ generated by the individual particles initially residing in the states specified by the quantum numbers $n$ is given by
\begin{equation}
I(t)=\sum_{n}f(n,t)I_{n}(t),
\label{eq:current}
\end{equation}
where $f(n,t)$ is  the aforementioned  nonequilibrium distribution function given by eq.~(\ref{Eq:NonEq}).\\
In fig.3(c) the time dependence of the total current is depicted for different topological charges. The results reveal that a higher topological charge $\ell_{\rm OAM}$ leads to a higher current. It is interesting to compare fig.3(c) with fig.2(b). Due to inertia related to the finite effective mass of the carriers, the voltage drop does not build instantaneously as the field is applied. The current in the wire however (cf. fig.3(c)), builds up yet much later in a (transport) time determined by the effective velocities of the tunneling, rescattering, and interfering current carrying states. Enhancing $\omega$ or the topological charge the current in the wire merges faster which is evidenced  by the results for $\ell_{\rm OAM}=-15$ and $\ell_{\rm OAM}=-20$.
The small oscillations in the current in the wire are related to the oscillations of $Q^{\rm In/Out}(t)$. The currents have a maximum around $t=1.7$~ps which is the time where the OAM light laser pulse is switched off. After that the currents decrease over the course of time which is associated with the weakening flux of the density out of the ring.\\
From the above it is evident that we can multiply the induced current by fabricating  well separated rings and attaching them in series to the wire. Each of the ring should then be irradiated with an OAM beam (e.g., by depositing on each ring an appropriate spiral phase plate that generates locally OAM light). Similarly, one may clamp the ring serially  between two wires and drive currents in both wires (in opposite directions) by OAM irradiations. The LDOS in such a case for the same pulse parameters and a topological charge $\ell_{\rm OAM}=-10$ is depicted in fig.4 endorsing this scenario which  can be viewed as a photogalvanic-type effect with the additional caveat that, via nanostructuring, we can steer the photogalvanic current both in direction and magnitude, as illustrated schematically in fig.4b.

Summarizing, On the basis of full-fledge quantum dynamical simulations we demonstrated that a focused laser pulse carrying orbital momentum irradiating a ring structure results in a radial centrifugal drift of the carrier which leads to a voltage drop between the inner and outer ring boundaries. Wiring the ring to a conductive straight channel and irradiating the ring with the OAM beam splashes a directed current in the wire whose direction, duration and strength is tunable by the pulse parameters such as the topological charge, the pulse width and the intensity. We also suggested possible ways to enhance the current and extract it in an effective way.

\section*{Methods}
For the rings we use a radial confinement  potential \cite{WCTanModel1996} $V(r)=\frac{a_1}{r^2}+a_2r^2-V_0$, where $r=\sqrt{x^2+y^2}$ and $V_0=2\sqrt{a_1a_2}$. The key parameters of this potential are as follows: the average radius of the ring is given by $r_0=(a_1/a_2)^{1/4}$, the width of the ring at the Fermi energy $\varepsilon_F$ is $\Delta r\approx\sqrt{8\varepsilon_F/m^*\omega_0^2}$, where $\omega_0=\sqrt{8a_2/m^*}$ and $m^*$ is the electron effective mass. For $r$ near $r_0$, the potential of the ring is parabolic: $V(r)\approx\frac{1}{2}m^*\omega_0^2(r-r_0)^2$. Taking $a_1\approx0$ then $V(r)$ describes a quantum dot.
To uncover the centrifugal photovoltaic effect we perform full numerical propagation on space-time grid of carriers wave function  as governed by
\begin{equation}
\small
i\hbar\partial_t\Psi_{l_0,m_0}(x,y,t)=\left\{-\frac{\hbar^2}{2m^*}\left(\partial_x^2+\partial_y^2\right) + \frac{ie\hbar}{2m^*}\left(2\vec{A}(t)\cdot\vec{\nabla}+\vec{\nabla}\cdot\vec{A}(t)\right) + \frac{e^2}{2m^*}A^2(t)+ V \right\}\Psi_{l_0,m_0}(x,y,t)
\label{Eq:TDSE}
\end{equation}
where  a gauge is used in which the scalar potential vanishes. In the plane $z = 0$, the OAM beam is taken as Laguerre-Gaussian (LG) mode with an on-axis phase singularity of the strength, i.e. vortex topological charge $\ell_{\rm OAM}$. In addition to $\ell_{\rm OAM}$ the LG modes are described by the radial index $p$ and the waist size $w_0$. Here we use the simplest form of the LG modes with $p=0$ in which case the intensity profile is ring-shaped around $z=0$ (the case $p\neq0$ adds no further qualitative information). The corresponding pulse vector potential in polar coordinates with $r(x,y)=\sqrt{x^2+y^2}$ and $\varphi(x,y)=\arctan{y/x}$ is given by
$A(r,\varphi,t)={\rm Re}\left\{\vec{e}A_0\left(\frac{\sqrt{2}r}{w_0}\right)^{|\ell_{\rm OAM}|} e^{ -r^2/w_0^2}e^{i(\ell_{\rm OAM}\varphi-\omega t)} \right\}$.  Due to computational limitations of the present full numerical time-propagation scheme we did not inspect larger or more complex structures. However, it is conceivable that the predicted effects are of a general nature  which are akin  both to the light vortex and the confinement effects. 
%
%
%
%
%\bibliography{v1}

%
%
%
%% Here is the endmatter stuff: Supplementary Info, etc.
%% Use \item's to separate, default label is "Acknowledgements"

\section*{Acknowledgements} We acknowledge financial support through the Deutsche Forschungsgemeinschaft under SPP 1840. Consultations on the experimental realization with J. Schilling, A. Sprafke, and R.B. Wehrspohn are gratefully acknowledged.
\section*{Author contributions}
 Both authors contributed equally to the  development of the idea, analysis of the results, and to writing the manuscript. J.W. conducted the numerical calculations.
\section*{Competing Interests} The authors declare that they have no competing financial interests.
\section*{Correspondence} Correspondence and requests for materials should be addressed to J.B. or J.W. (email: Jamal.Berakdar@physik.uni-halle.de or
Jonas.Waetzel@physik.uni-halle.de).
% \end{addendum}
%
%%
%% TABLES
%%
%% If there are any tables, put them here.
%%
%
%
%
%
%
\end{document}